\documentclass[aps,prl,twocolumn,floats,showpacs]{revtex4}
\usepackage{graphicx}
\usepackage{amsmath}
\usepackage{amssymb}
\usepackage{color}

\begin{document}
\title{Control of molecular rotation with a chiral train of ultrashort pulses}

\date{\today}
\author{S.~Zhdanovich$^{1}$,  A.A.~Milner$^{1}$, C.~Bloomquist$^{1}$, J.~Flo{\ss}$^{2}$, I.Sh.~Averbukh$^{2}$, J.W.~Hepburn$^{1}$, and V.~ Milner$^{1}$}
\affiliation{$^{1}$Department of  Physics \& Astronomy and The Laboratory for Advanced Spectroscopy and Imaging Research (LASIR), The University of British Columbia, Vancouver, Canada \\
$^{2}$Department of Chemical Physics, The Weizmann Institute of Science, Rehovot, Israel}

\begin{abstract}{Trains of ultrashort laser pulses separated by the time of rotational revival (typically, tens of picoseconds) have been exploited for creating ensembles of aligned molecules. In this work we introduce a \textit{chiral pulse train} - a sequence of linearly polarized pulses with the polarization direction rotating from pulse to pulse by a controllable angle. The chirality of such a train, expressed through the period and direction of its polarization rotation, is used as a new control parameter for achieving selectivity and directionality of laser-induced rotational excitation. The method employs chiral trains with a large number of pulses separated on the time scale much shorter than the rotational revival (a few hundred femtosecond), enabling the use of conventional pulse shapers.}
\end{abstract}


\maketitle

Control of molecular rotation with strong non-resonant laser fields has become a powerful tool for creating ensembles of aligned \cite{Friedrich1995, Stapelfeldt2003, Averbukh2001, Vrakking2001}, oriented \cite{Averbukh2001, Rost1992, Vrakking1997} and spinning molecules \cite{Karczmarek1999, Kitano2009, Fleischer2009, Hoque2011}. Numerous applications of rotational control in molecular systems include control of chemical reactions \cite{Stapelfeldt2003, ShapiroBrumerBook}, deflection of neutral molecules by external fields \cite{Gershnabel2010, Stapelfeldt1997, Purcell2009}, high harmonic generation \cite{Itatani2005, Wagner2007}, and control of molecular collisions with atoms \cite{Tilford2004} and surfaces \cite{Tenner1991, Kuipers1988, Greeley1995, Zare1998, Shreenivas2010}. Alignment of molecular axes has been implemented with transform-limited and shaped laser pulses using various approaches (see, e.g. \cite{Friedrich1995, Stapelfeldt2003, Underwood05, Lee06, Daems05, Holmegaard09}). Increasing the degree of molecular alignment has been achieved by employing a sequence of laser pulses (a ``pulse train''), separated by the time of rotational revival \cite{Leibscher2003, Cryan2009, Zhao2011}. Such timing of pulses in the train ensures that the accumulative effect of consecutive pulses is equally efficient for all molecules in the ensemble, regardless of their angular momentum.

Creating molecular ensembles with a preferred direction of rotation has been reported in both adiabatic (``optical centrifuge'') \cite{Karczmarek1999, Villeneuve2000, Vitanov2004, Yuan2011, Cryan2011} and non-adiabatic \cite{Fleischer2009, Kitano2009} regimes of excitation using a pair of laser pulses with different polarization. In this work we demonstrate an alternative way of exciting uni-directional rotational motion with a ``chiral pulse train'', in which the polarization of the excitation field rotates from pulse to pulse by a controllable angle, in either clockwise or counter-clockwise direction. The time delay between the pulses is much shorter than the revival period. We show that by varying the train parameters, one can achieve selectivity in the rotational excitation and control its directionality. By matching the rotational period of the field polarization to the period of molecular rotation, molecules in a particular angular momentum state can be excited more efficiently than others. The chirality and the period of the pulse train define the direction of the molecular rotation, which is detected by multi-photon ionization with circularly polarized light.

\begin{figure}
\centering
    \includegraphics[width=0.75\columnwidth]{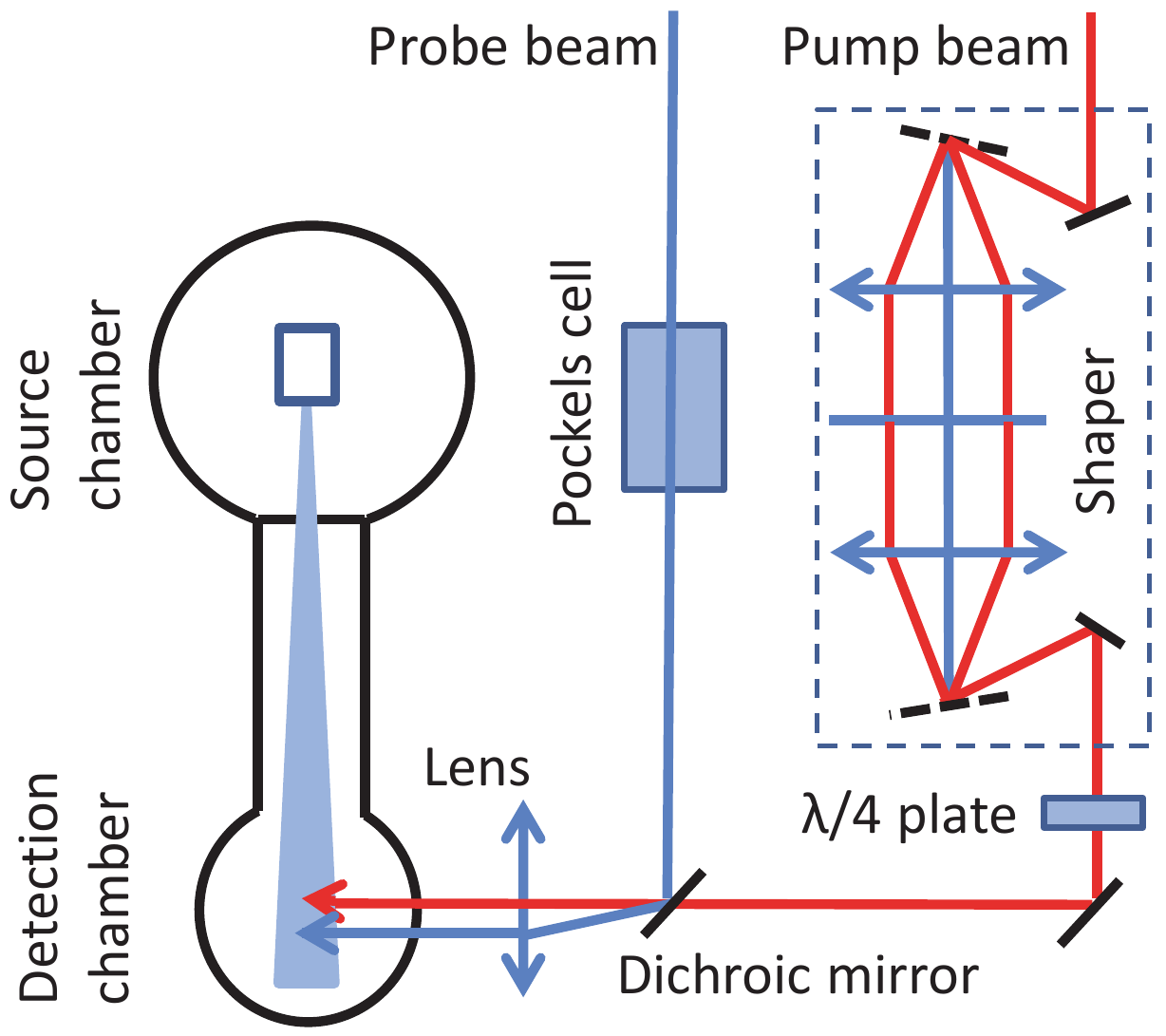}
     \caption{Diagram of the experimental setup. The pump pulse originated from a Ti:Sapphire amplifier is shaped in a home build shaper and focused on a cold jet of Oxygen in the detection chamber. The probe pulse produced by a tunable dye laser is combined with the pump on a dichroic mirror. Circular polarization of probe pulses is switched from clockwise to counter-clockwise by a Pockels cell and a $\lambda/4$ plate. The probe pulse is delayed with respect to the pump pulse and its focus is shifted downstream.}
  \vskip -.1truein
  \label{FigSetup}
\end{figure}

Our experimental setup is depicted in Fig.\ref{FigSetup}. Cold oxygen molecules are produced by a supersonic expansion in a vacuum chamber. The rotational temperature of the molecules in the jet is 7-9 K. The molecular jet enters the detection chamber through a 1-mm diameter skimmer. Pump pulses for rotational excitation are produced by a femtosecond Ti:Sapphire amplifier (120 fs, 2 mJ at 800 nm and 1 KHz repetition rate) and a home made pulse shaper based on a liquid crystal spatial light modulator (SLM). The pulse length is much shorter than the rotational period of Oxygen in the lower rotational states considered in this work. The rotational distribution is probed by narrowband nanosecond pulses generated by a tunable dye laser (3 mJ at 285 nm and 10 Hz repetition rate). The molecules are ionized via a ``2 + 1'' resonance enhanced multi-photon ionization (REMPI) process, with a two-photon resonant transition $C^{3}\Pi _{g}(v'=2) \leftarrow X^{3}\Sigma _{g}^{-}(v''=0)$ of O$_{2}$. The total ion signal is detected by a multi-channel plate in a standard time-of-flight detection apparatus. Pump and probe beams are combined on a dichroic mirror and focused on the molecular jet with a 150 mm focal length lens. A Pockels cell is used to alternate circular polarization of the consecutive probe pulses between left and right, allowing precise detection of the induced circular anisotropy. In order to reduce the ionization background from pump pulses, the position of the pump beam is shifted by a few hundred $\mu$m upstream with respect to the probe beam, while the time delay between the pulses is set to let the excited molecules reach the probe focal spot.  We estimate the intensity of pump pulses of the order of $10^{12}$ W/cm$^2$, which results in a dimensionless spatially-averaged ``rotational kick strength'' of $P\approx7$. The latter parameter characterizes the amount of angular momentum (in units of $\hbar$) transferred from the laser field to the molecule \cite{Averbukh2001, Leibscher2003}.

\begin{figure}
\centering
    \includegraphics[width=0.9\columnwidth]{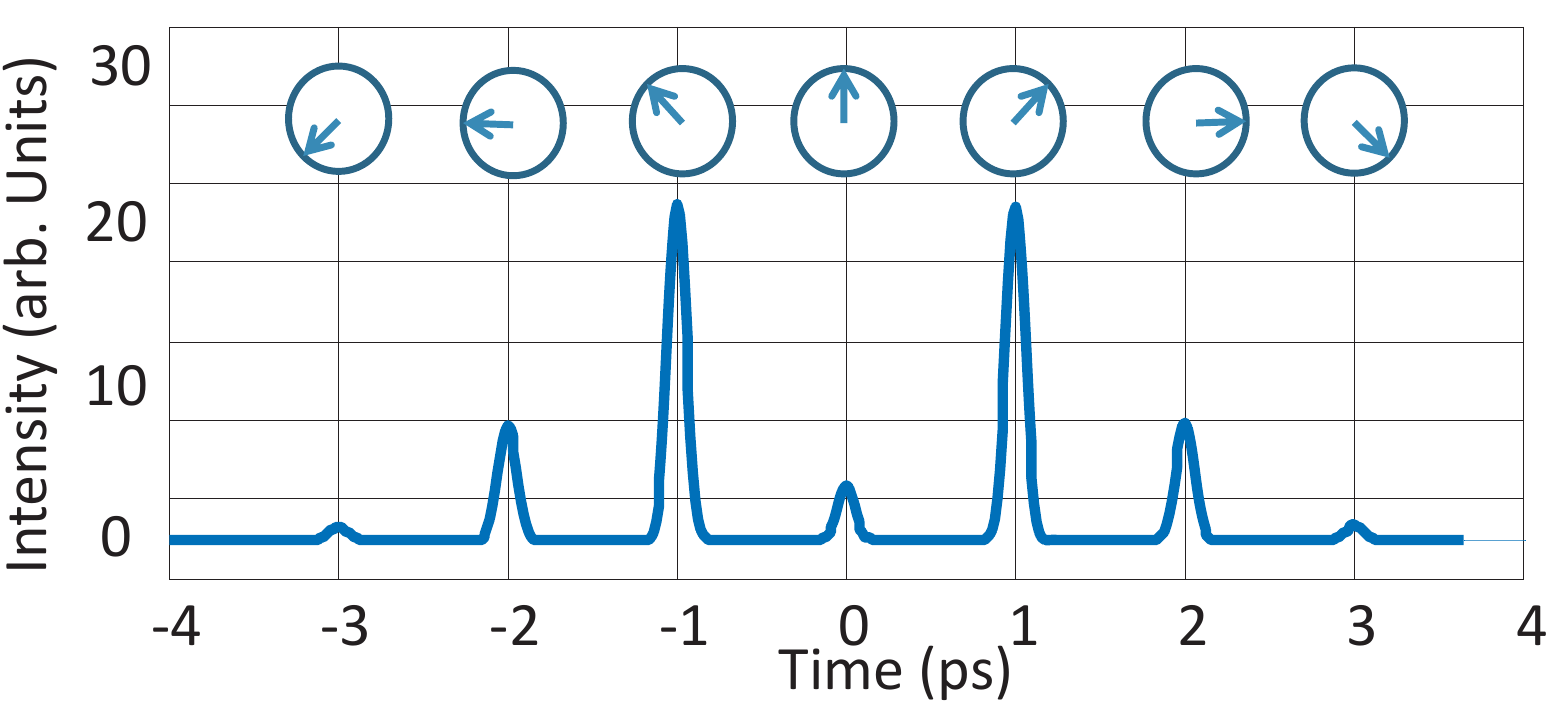}
     \caption{Example of an intensity envelope and polarization of a chiral pulse train implemented in this work. The train parameters are: $A=2$, $\delta=\pi/4$, and $\tau =$ 1 ps. Polarization of each pulse, rotating from pulse to pulse with a period of 8 ps, is shown in the circles above.}
  \vskip -.1truein
  \label{FigPulseTrain}
\end{figure}

An example of a chiral pulse train, used in this work, is shown in Fig.\ref{FigPulseTrain}. It is produced by a spectral pulse shaper implemented in a standard $4f$ geometry with a double-layer SLM in its Fourier plane \cite{Weiner2000}. The two shaper masks control the spectral phase $\varphi_{1,2}(\omega )$ of the two polarization components of an input pulse along the two orthogonal axes of the shaper, $\hat{e}_1$ and $\hat{e}_2$. If $\varphi_{1,2}(\omega )=A\sin[(\omega-\omega_{0})\tau +\delta_{1,2}]$, where $\omega_0$ is the optical carrier frequency, $A$ is the modulation amplitude, $\tau $ is the train period and $\delta _{1,2}$ are two arbitrary angles, the resulting field in the time domain is:
\begin{equation}
\label{EqRotPolPulseTrain}
E(t)=\sum_{i=1,2}{\hat{e}_i (\hat{e}_i \cdot \hat{e}_\text{in})  \sum_{n=-\infty}^{\infty}{J_n(A)\varepsilon(t+n\tau ) \cos[\omega_0t+n\delta_{i}]}},
\end{equation}
where $\varepsilon(t)$ is the electric field envelope of the original pulse polarized along $\hat{e}_\text{in}$. Eq.\ref{EqRotPolPulseTrain} describes a train of elliptically polarized pulses, with the polarization ellipticity of the $n$-th pulse defined by the phase difference $n(\delta _{1}-\delta _{2})$. A quarter-wave plate, oriented along $\hat{e}_\text{in}$, converts this elliptical polarization back to linear, rotated by an angle $n(\delta _{1}-\delta _{2})/2$ with respect to the input polarization. By choosing $\delta _{1}= - \delta _{2} = \delta $, we can create a pulse train with the polarization of each pulse rotated with respect to the previous one by angle $\delta $. The period of the polarization rotation is $T_{p}=2\pi \tau /\delta $. In Fig.\ref{FigPulseTrain} an example of a pulse train is shown for the modulation amplitude $A=2$, train period $\tau =$ 1 ps, and polarization rotation period $T_{p}=$ 8 ps. We experimentally characterized the train by polarization sensitive cross-correlation technique \cite{OurLongPaper}.

\begin{figure}
\centering
    \includegraphics[width=1.0\columnwidth]{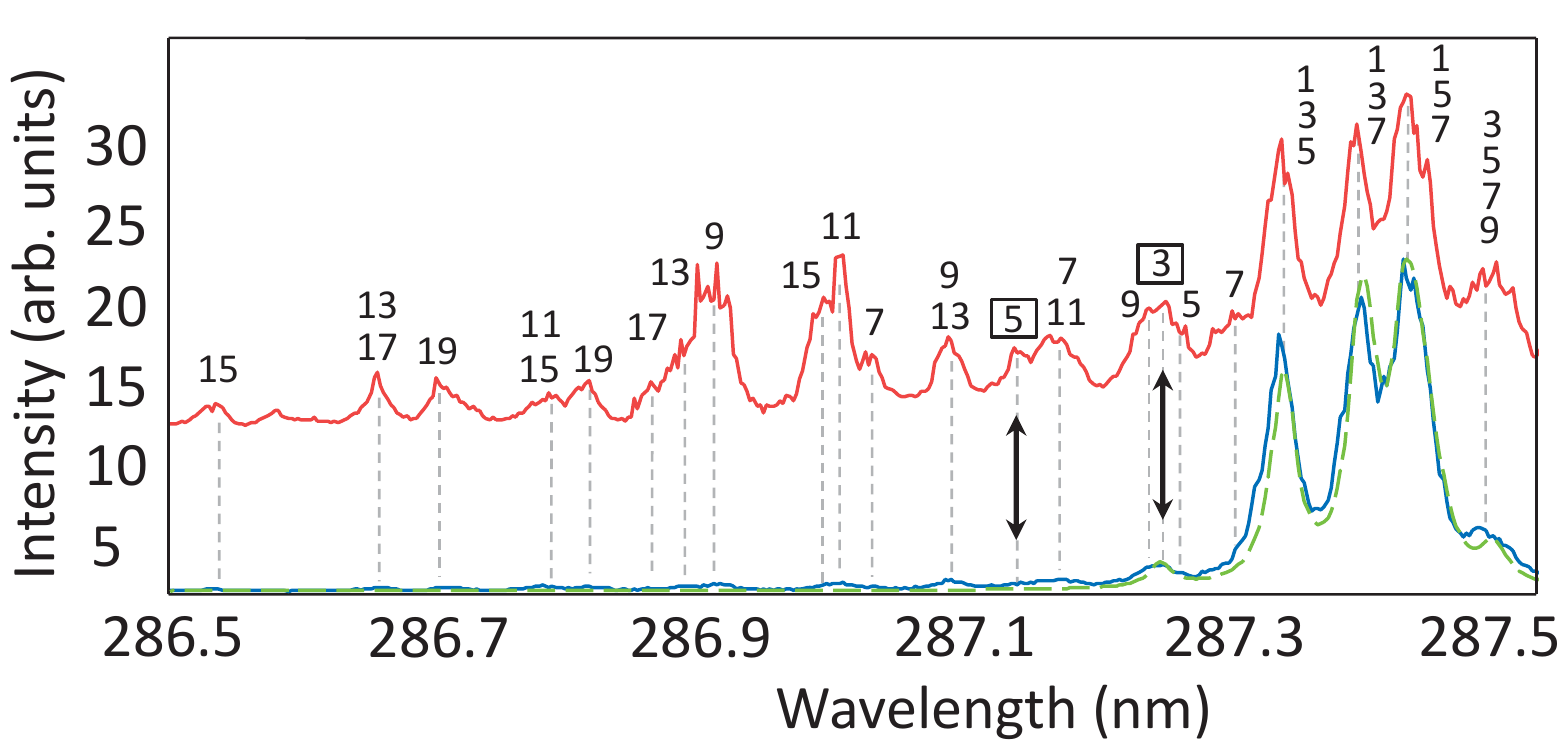}
     \caption{REMPI spectrum of jet-cooled Oxygen: experimental results (lower solid curve) and calculations (lower dashed line). Upper solid curve shows the spectrum of rotationally excited molecules (shifted up for clarity). Dashed vertical lines indicate the frequencies of transitions originated from a certain rotational state, labeled with the corresponding $N''$ number. Thick vertical arrows point at the two peaks which correspond mostly to $N''$=3 (right), and $N''$=5 (left) initial rotational states.}
  \vskip -.1truein
  \label{FigREMPI}
\end{figure}

The observed REMPI spectrum of jet-cooled Oxygen is shown in Fig.\ref{FigREMPI}, together with a calculation (carried out similarly to \cite{Mizushima54}) for the rotational temperature of 8 K. The majority of molecules is in the lowest rotational state $N''=1$. Application of a pump pulse train changes the spectrum dramatically by increasing the intensity of lines originated from $N''>1$. REMPI spectrum of the excited molecules, obtained with the maximum available pulse energy of 300 $\mu $J, shows populated peaks for the rotational number as high as $N''=19$.  Unfortunately, different rotational transitions in O$_{2}$ overlap with one another and it is generally difficult to have a REMPI line corresponding to a single initial state (see line assignments in Fig.\ref{FigREMPI}). We use two spectral lines at 287.25 nm and 287.14 nm, for which the majority of ions originates from $N''=3$ and $N''=5$ rotational states, respectively. Even though the individual lines are not resolved, the relative contribution of the neighboring states to the total ion signal in both cases is estimated as $<18$\%.

\begin{figure*}
\begin{center}
    \begin{minipage}[t]{1\linewidth}
    \includegraphics[width=1\columnwidth]{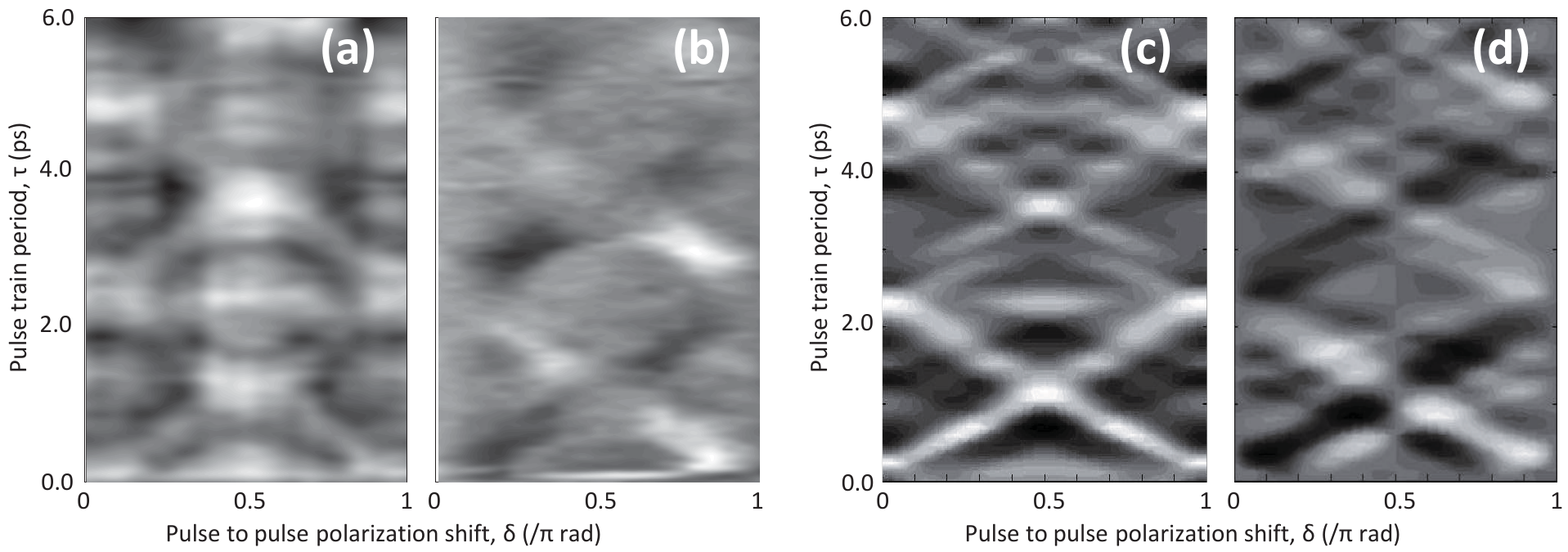}
    \caption{Total excitation efficiency (a,c) and circular anisotropy (b,d) of Oxygen molecules excited to $N''=3$ rotational state. (a,b) - experimental data, (c,d) - numerical calculations. Each experimental data point represents an average over 150 probe pulses. Color coding: (a) black: $S = 0$, white: $S = 1$ (arb.units); (b) black: $\epsilon = -0.2$, white: $\epsilon = +0.2$; (c) black $S=0.1$, white $S=0.6$; (d) black: $\epsilon = -0.7$, white: $\epsilon = +0.7$;}
    \label{FigN3Results}
    \end{minipage}\hfill
  \end{center}
\end{figure*}
\begin{figure*}
\begin{center}
    \begin{minipage}[t]{1\linewidth}
    \includegraphics[width=1\columnwidth]{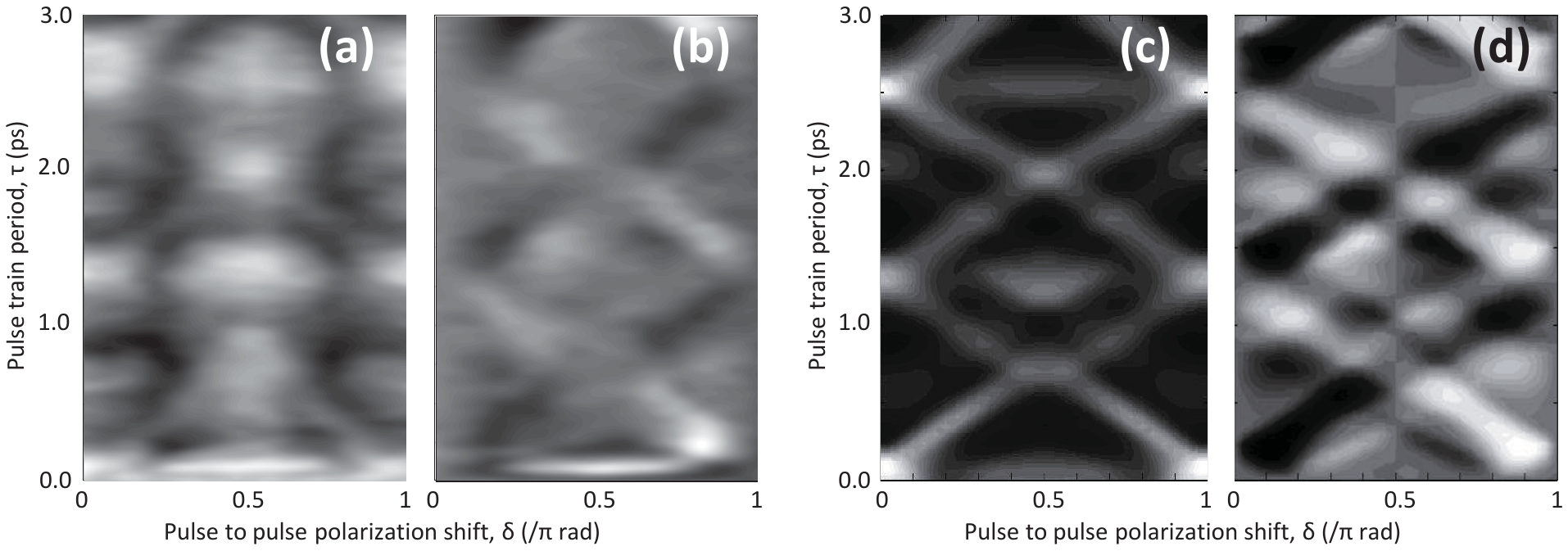}
    \caption{Same as Fig.\ref{FigN3Results}, but for molecular excitation to $N''=5$ rotational state. Color coding: (a) black: $S = 0$, white: $S = 1$ (arb.units); (b) black: $\epsilon = -0.4$, white: $\epsilon = +0.4$; (c) black: $S = 0$, white: $S = 0.35$; (d) black: $\epsilon = -0.8$, white: $\epsilon = +0.8$;}
    \label{FigN5Results}
    \end{minipage}\hfill
  \end{center}
\end{figure*}

Our main results are shown in Figs.\ref{FigN3Results} and \ref{FigN5Results}. With the probe wavelength set to 287.25 nm ($N''=3$) and 287.14 nm ($N''=5$), we vary $\delta$ and $\tau$ while keeping the pulse train envelope and energy (140 $\mu $J) constant. To measure the directionality of molecular rotation with high accuracy and low susceptibility to the power fluctuations of the nanosecond dye laser, we modulate the polarization of probe pulses at 10 Hz with a Pockels cell and record the total ionization signal for left and right circularly polarized probe, $S_{L}$ and $S_{R}$, respectively. The sum of the two signals, $S= S_L+S_R$, represents the average total efficiency of exciting the molecules to $N''=3$ or 5. Panel (a) shows the experimentally detected value of $S$ as a function of $\delta$ and $\tau$. We define the degree of rotational directionality as $\epsilon  =(S_L-S_R)/(S_L+S_R)$, and show it in panel (b).

The experimental results are compared with a theoretical analysis of the population distribution of various rotational states of oxygen excited by the chiral train. In our model, the pulses were considered as $\delta$-kicks (impulsive approximation), and the molecular wavefunction was expanded in the Hund's case (b) basis. The non-perturbative modification of the expansion coefficients due to the interaction with every pulse was determined with the help of the numerical procedure described in \cite{Fleischer2009}. Thermal averaging over initial molecular states was done to account for thermal effects. As the observables related to the measured signals, we calculated  $Q_L$ ($Q_R$) - the total population of the $N$-states with positive (negative) projection $M_J$ of the total angular momentum $\mathbf{J}$ onto the propagation direction of the chiral pulse train. The population of the $M_J=0$ state was equally divided between $Q_L$ and $Q_R$. The sum of the two signals, and the normalized directionality $(Q_L-Q_R)/(Q_L+Q_R)$ are shown in panels (c) and (d) of Fig.\ref{FigN3Results}, and should be compared with the experimental results of Fig.\ref{FigN3Results}(a) and (b), respectively.

On plot (a) of Fig.\ref{FigN3Results} one can see that for $\delta=0, \pi$ (non-rotating polarization) the total signal exhibits well pronounced maxima at $\tau\approx 2400$ fs and  $\tau\approx 4700$ fs. As expected, these times correspond to one and two periods of ``rotation'' for $N''=3$, defined as $T_{N=3}=h/(E_{N=3}-E_{N=1})=2340$ fs. Of course, the picture of classical rotation is not applicable for such low rotational numbers, and the ``period of rotation'' simply means the evolution period of a rotational wavepacket consisting of only two states, $N''=1$ and $N''=3$.

Fig.\ref{FigN3Results}(a) shows an ``X'' pattern with clear diagonal lines. Their slope defines a constant period of polarization rotation in the chiral train, $T_{p}=2\pi\tau/\delta $. Diagonals with a positive slope, e.g. from point $(\tau=0, \delta=0)$ to point $(\tau=2400, \delta=\pi)$, correspond to the polarization rotating clockwise, while a negative-slope diagonal corresponds to counter-clockwise polarization rotation. The population of the corresponding state is clearly higher along these diagonal lines, reflecting higher degree of rotational excitation by a pulse train whose polarization is rotating in sync with the molecules. Note that the lines of enhanced excitation correspond to the train polarization rotating twice slower than the molecules, e.g. $T_{p}=2T_{N=3}$. The effect is reproduced in the numerical calculations shown in panel (c), and can be attributed to the inversion symmetry of the oxygen molecule.

The directionality of molecular rotation along the ``resonant'' diagonals of plot (a) is clearly confirmed by our measurement of the circular anisotropy of the excited rotation, $\epsilon $, shown in plot (b) of Figs.\ref{FigN3Results} and \ref{FigN5Results}. Indeed, positive-slope diagonals correspond to $\epsilon <0$ (clockwise rotation), whereas those with a negative slope show $\epsilon >0$ (counter-clockwise rotation). Following the diagonal lines in plot (a), one can notice that the signal is higher at $\delta=0, \pi/2$ and $\pi$ (middle and end points) in comparison with the intermediate values of $\delta $. This is the result of a bi-directional rotational excitation in those points, where the pulse train is resonant with both clockwise and counter-clockwise rotating molecules, and the total number of excited molecules is therefore higher. This conclusion is evident in plot (b), where no circular anisotropy ($\epsilon =0$) is observed around $\delta=0, \pi/2, \pi$. The described effect is confirmed by the numerical results, shown in panel (d).

An interesting increase in the directionality of the excited rotation is observed for low train period ($\tau=100$ fs) and polarization rotation angle $\delta \approx \pi/2$. In this case, pulses of the chiral train are overlapping in time, forming a single pulse with gradually rotating polarization. This is exactly the centrifuge field of \cite{Karczmarek1999, Villeneuve2000, Yuan2011, Cryan2011}. Even though our centrifuge pulse is of much lower strength and duration than that needed for spinning molecules to high angular frequencies, its effect on low rotational state is quite evident for both $N''=3$ and $N''=5$ (Figs.\ref{FigN3Results}(b) and \ref{FigN5Results}(b), respectively). The centrifuge effect is not observed in the numerical calculations, because the latter assume an infinitely short duration of pulses in the chiral pulse train.

In summary, we have proposed and implemented a new method of exciting uni-directional molecular rotation with polarization-shaped femtosecond laser pulses. The technique of generating a sequence of pulses with field polarization rotating from pulse to pulse by a predefined constant angle - a chiral pulse train, has been demonstrated and utilized for the rotational control of molecular Oxygen. We show that tuning the parameters of the chiral pulse train enables one to control both the frequency of excited molecular rotation and its directionality.

\begin{acknowledgements}
The authors would like to thank Guillaume Bussiere for the help with the REMPI setup. This work has been supported by the CFI, BCKDF and NSERC. IA and JF acknowledge support from  the Israel Science Foundation. SZ is a recipient of an Alexander Graham Bell scholarship from NSERC. JF is a recipient of a fellowship from Minerva Foundation.
\end{acknowledgements}


\end{document}